# Characterization Of any Non-linear Boolean function Using A Set of Linear Operators

[1]**Sudhakar Sahoo**, [2]**Pabitra Pal Choudhury**, [3]**Mithun Chakraborty**
[1,2]Applied Statistics Unit, Indian Statistical Institute, Kolkata, 700108, INDIA
Email: sudhakar.sahoo@gmail.com, pabitrapalchoudhury@gmail.com
[3]Dept.of Electronics and Telecomm. Engg., Jadavpur University, Kolkata-700032
Email: mithun.chakra108@gmail.com

**Abstract: -** Global dynamics of a non-linear Cellular Automata is, in general irregular, asymmetric and unpredictable as opposed to that of a linear CA, which is highly systematic and tractable. In the past efforts have been made to systematize non-linear CA evolutions in the light of Boolean derivatives and Jacobian Matrices. In this paper two different efforts have been made: first we try to systematize non-linear CA evolution in the light of deviant states and non-deviant states. For all the non-deviant states the nearest linear rule matrix is applicable where as for the deviant states we have a set of other matrices. Second using algebraic manipulation, an efficient algorithm is proposed by which every Non-linear Boolean function can be characterized by a sequence of binary matrices.

## 1. Introduction

Boolean logic was named after George Boole, who first defined an algebraic system of logic in the mid 19th century. Boolean logic has many applications in electronics, computer hardware and software, and is the base of digital electronics. On the other hand, Cellular Automata (CA) introduced by J. von Neumann [6] is a suitable tool to handle Complex systems. CA rules have many real life applications in almost all area of science like Physics, Chemistry, Mathematics, Biology, Engineering, Finance etc. A connection can be made between CA rules in different dimensions with n variables Boolean functions [4, 7]. Out of $2^{2^n}$ number of Boolean functions we have $2^n$ are linear and the rest are non-linear. In this way we get linear CAs and non-linear CAs [4].

The *dynamic behavior* of any CA is visualized and studied in terms of either its *space-time pattern* or its *basin-of-attraction* field [12]. The latter is essentially a *graph*, which may or may not consist of disjoint sub graphs, and is commonly referred to as the *State Transition Diagram* or, in short, the STD of the CA. All linear CA are symmetric in representation and have a linear handle. A single matrix can represent a linear function for any input string [1, 2]. Also its STD's are well structured and symmetric in nature. The Non-linear functions on the other hand are non-uniform and asymmetrical in representation in the state transition diagrams. No single matrix can represent a Non linear function for any input string. Thus any linear Uniform Cellular Automata (UCA) STD may be taken as a standard for comparison because all its essential features bear simple and well-known relationships with the fundamental properties such as rank, nullity, determinant etc. of the *state transition matrix* or *transformation matrix*, denoted by *T*, of the corresponding linear CA rule. With this in mind, we have made an attempt at the *relative characterization* of a particular set of non-linear UCA STDs by first identifying the *nearest linear rule* of each such non-linear rule, then considering the STD of the said nearest linear CA rule as a *linear model* for the non-linear STD concerned and finally determining the nature and extent of departure of this non-linear STD from the said linear model. A set of deviant states and non-deviant states between these rules are computed. For all the non-deviant states the nearest linear rule matrix is applicable where as for the deviant states we have a set of other matrices. Second, an efficient algorithm is proposed that identifies a set of minimum number of matrices as a representative of any arbitrary Non linear Boolean function.

In section 2, some preliminary discussions on both Boolean functions and Cellular Automata are discussed. In [2] Boolean functions are classified and sub classified according to their degree of non-linearity and also the position of bit mismatch some of the ideas are included in section 3. Section 4 introduces the concept of deviant and non-deviant states. For non-linear CA, the STD and a set of matrices can be computed from the STD of nearest linear CA. An efficient algorithm is proposed in section 5 by which every Non-linear Boolean function can be characterized by a sequence of binary matrices. Section 6 concludes the paper.

## 2. Basic concepts

### 2.1 Boolean Functions

A Boolean function $f(x_1, x_2, ..., x_n)$ on n-variables is defined as a mapping from $\{0,1\}^n$ into $\{0,1\}$. It is also interpreted as the output column of its truth tables f i.e. a binary string of length $2^n$. For n variables the number of Boolean Functions is $2^{2^n}$. Boolean Functions can be expressed using its Algebraic Normal Form (ANF)[2, 3, 8, 9, 10, 11] i.e. in the form of XOR operations but have been referred to by their Wolfram numbers [4]. Rule *W* may be sometimes denoted as $f_W$.

For three variables, the number of Boolean functions is $2^{2^3} = 256$. The three variable ones can also be represented in DNF and ANF. In DNF, $f_1 = x'.y'.z'$ and in ANF $f_1 = (x \oplus 1)(y \oplus 1)(z \oplus 1) = xyz \oplus xy \oplus xz \oplus x \oplus yz \oplus y \oplus z \oplus 1$. Similarly Rule 150 can be denoted as $f_{150} = x \oplus y \oplus z$ *etc*. If the Wolfram's number of a rule is even (odd), its ANF number is also even (odd), hence, without loss of generality, a rule may be referred to as "even-numbered" rule or "odd-numbered", as the case may be. Thus Rule 10 is an *even* rule while Rule 57 is an *odd* rule.

The Boolean functions are of two types: Linear and Non-linear. Consider a Boolean function in ANF form: $f(x_1, x_2, ..., x_n) = a_0 \oplus a_i x_i \oplus a_{i,j} x_i x_j \oplus ... \oplus a_{1,2,...n} x_1 x_2 ... x_n$ for $1 \leq i \leq n$ and $1 \leq i < j \leq n$

The number of variables in the highest product term with non zero coefficients is the algebraic degree. A function of degree at most one is called Affine function. An Affine function with constant term equal to zero is called a linear function. Non-linearity of an n variable Boolean function is the distance from the set of all n variable Affine functions. $f_0, f_{170}, f_{204}, f_{102}, f_{240}, f_{90}, f_{60}, f_{150}$ are the linear functions in 3 variables. The rest of the 248 functions are called Non-linear functions.

### 2.2. Terminology and notation pertaining to one-dimensional cellular automata

In this paper, we shall restrict ourselves to the study of a one-dimensional, binary cellular automaton (CA) of *n* cells (i.e. *n* bits) $x_1, x_2, ..., x_n$, with local architecture[3]. The *global state* or simply *state* of a CA at any time-instant *t* is represented as a vector $X^t = (x_1^t, x_2^t, ..., x_n^t)$ where $x_i^t$ denotes the bit in the $i^{th}$ cell $x_i$ at time-instant *t*. However, instead of expressing a state as a bit-string, we shall frequently represent it by the decimal equivalent of the *n*-bit string with $x_1$ as the Most Significant Bit; e.g. for a 4-bit CA, the state *1011* may be referred to as state **11** ($=1 \times 2^0 + 1 \times 2^1 + 0 \times 2^2 + 1 \times 2^3$).

The bit in the $i^{th}$ cell at the "ne*x*t" time-instant *t+1* is given by a *local mapping* denoted by $f^i$, say, which takes as its argument a vector of the bits (in proper order) at time-instant *t* in the cells of a certain pre-defined *neighborhood* (of size *p*, say) of the $i^{th}$ cell. Thus, the size of the neighborhood is taken to be the same for each cell and may also be called the 'number of variables' (which $f^i$ takes as inputs).

**Null boundary (*NB*) :** The left neighbor of $x_1$ and the right neighbor of $x_n$ are taken as 0 each.
**Periodic boundary (*PB*) :** $x_n$ is taken as the left neighbor of $x_1$ and $x_1$ as the right neighbor of $x_n$.

A CA may be represented as a string of the rules applied to the cells in proper order, along with a specification of the boundary conditions. e.g. **<103, 234, 90, 0>NB** refers to the CA ($x_1, x_2, x_3, x_4$) where $x_1^{t+1} = f_{103}(0, x_1^t, x_2^t)$; $x_2^{t+1} = f_{234}(x_1^t, x_2^t, x_3^t)$; $x_3^{t+1} = f_{90}(x_2^t, x_3^t, x_4^t)$; $x_4^{t+1} = f_0(x_3^t, x_4^t, 0)$.

If the "present state" of an *n*-bit CA (at time *t*) is $X^t$, its "next state" (at time *t+1*), denoted by $X^{t+1}$, is in general given by the *global mapping* $F(X^t) = (f^1(lb^t, x_1^t, x_2^t), f^2(x_1^t, x_2^t, x_3^t), ..., f^n(x_{n-1}^t, x_n^t, rb^t))$, where *lb* and *rb* denote respectively the left boundary of $x_1$ and right boundary of $x_n$.

If the rule applied to each cell of a CA is a linear Boolean function, the CA will be called a **Linear Cellular Automaton,** otherwise a **Non-linear Cellular Automaton,** e.g.<0, 60, 60, 204>NB is a linear CA while <31,31,31,31>NB and <60,90,87,123>PB are non-linear CAs.

If the same Boolean function (rule) determines the "ne*x*t" bit in each cell of a CA, the CA will be called a **Uniform Cellular Automaton (UCA)**, otherwise it will be called a **Hybrid Cellular Automaton (HCA)**, e.g.<135, 135, 135, 135>PB is a UCA, <0, 60, 72, 72>NB is a HCA.

For a UCA, the Boolean function applied to each cell will be called the **rule of the CA**. So for a UCA, we can obviously drop the superscript '*i*' from the local mapping $f^i$ and simply denote it as *f*. e.g. for the 4-bit CA <230, 230, 230, 230>PB, the rule of the CA is Rule 230 and the CA will be called the "Rule 230 CA" of 4 bits with periodic boundary conditions. For our purpose, we shall be mostly interested in *elementary CA* defined by Wolfram [3] to be one-dimensional binary CA with a symmetrical neighborhood of size *p* = 3 for each cell so that $x_i^{t+1} = f^i(x_{i-1}^t, x_i^t, x_{i+1}^t)$, *i* = 2,3,......, *n*-1.

## 2.3. Boolean derivatives and Jacobian Matrix

The *first-order partial Boolean derivative* [1] of a Boolean function $f(x_1, x_2, \ldots, x_n)$ with respect to $x_j$, $j = 1, 2, \ldots, n$ is defined as $\partial f / \partial x_j = f(x_1, x_2, \ldots, x_j, \ldots, x_n) \oplus f(x_1, x_2, \ldots, \bar{x}_j, \ldots, x_n)$ Where $\bar{x}_j$ is the Boolean complement of $x_j$.

The *gradient* of a Boolean function $f(x_1, x_2, \ldots, x_n)$, denoted by grad($f$) is defined as the vector of the $n$ first-order partial Boolean derivatives of the function with respect to the $n$ input variables *in the proper order*, i.e. grad($f$) = [ $\partial f / \partial x_1 \quad \partial f / \partial x_2 \ldots \ldots \partial f / \partial x_n$ ]

The Jacobian matrix of an n-bit one-dimensional CA is defined as an $n \times n$ binary matrix, denoted by J, whose (i, j)th entry is $J_{ij} = \partial f^i / \partial x_j^t \; \forall \, i \in \{1,2,3,\ldots,n\}, \; \forall \, j \in \{1,2,3,\ldots,n\}$. Under the assumption $p = 3$, the Jacobian matrix is a tri-diagonal matrix, except for the two *off-diagonal corner elements* in the periodic-boundary case.

For all linear CA rules and equivalently for all linear Boolean functions there exist an $(n \times n)$ matrix $A$ which when multiplied by an $n$-bit string $x = (x_1, x_2, \ldots x_n)$ except $x = (0, 0, \ldots, 0)$ gives its corresponding $n$-bit output $y = (y_1, y_2, \ldots y_n)$. That is if the rule is linear then $y = (Ax)$, Where $x = (x_1, x_2, \ldots x_n)$ and $y = (y_1, y_2, \ldots y_n)$. This matrix is same as its Jacobian matrix [1, 2]. This type of matrices can be constructed for other linear rules in one dimension and therefore a single matrix of order $(n \times n)$ exists for every linear rule applied to arbitrary n-bit string. But for non-linear rules, is there any matrix or sequence of matrices exists for all possible n-bit strings. Our next discussion gives the answer for this.

## 3. Studies on the H.D's between Boolean functions

The *Hamming distance* (abbreviated as H.D. throughout this paper) between any two bit sequences *of equal length* is defined as the number of positions at which the bits differ in the two sequences.

The H.D. between two Boolean functions of $n$ binary variables is defined as the H.D. between the $n$-bit binary equivalents of the rule numbers according to Wolfram's labeling convention [3]. For *e*xample, let us take two Boolean functions of three variables viz. Rule 34 and Rule 225. Their 8-bit binary representations are **00**100**0**1**0** and **11**100**00**1 respectively. Clearly, these two strings differ from each other at 4 bit-positions. Hence, the H.D. of Rule 34 from Rule 225 is **4**. Equivalently, the H.D. between two rules $f_1$ and $f_2$ is given by the *weight* of the sum mod 2 of these two rules (viz. $f_1 \oplus f_2$), the *weight* of a Boolean function being defined as the number of '1's in the output column of its Truth Table. It is worthwhile to mention here that the *minimum H.D.* between a Boolean function *f* and the set of all affine functions is called the *degree of non-linearity* of *f*.

## 3.1. Theorems on Boolean functions and Jacobian matrices

**Theorem 3.1:** If the H.D. of an *n*-variable Boolean function *f* from another rule *g* is *m*, then the H.D. of the complement of *f* from the same rule *g* is $(2^n - m)$.

**Corollary 1:** For any non-linear rule of *n* variables, there exists at least one affine rule of *n* variables such that the H.D. between the two is smaller than or equal to $2^{n-1}$.

**Theorem 3.2:** The Jacobian matrices of two UCAs of the same size, with the same boundary conditions but with different rules, are identical *if and only if* the rule of one of the CAs is the Boolean complement of that of the other CA.

*Corollaries:*
All the corollaries to theorem 3.2 are stated in terms of CA with *p*=3 (section 2.2), although they are fairly general.

**(i)** Irrespective of the number of bits in the CA, is possible to have $2^{2^3} = 256$ different UCAs with either type of boundary condition (null/periodic) – since there are 256 different Boolean functions of 3 variables – but there are only 256/2 = *128* distinct Jacobian matrices (for given boundary conditions), each characterizing a pair of Boolean functions which are logical complements of each other; e.g. the complement of Rule 30 is Rule 225(=255 – 30), hence each of the UCAs <30, 30, 30, 30>PB and <225, 225, 225, 225>PB has the Jacobian matrix

$$J_{30}\,|_{PB} = J_{225}\,|_{PB} = \begin{bmatrix} \bar{x}_2 & \bar{x}_1 & 0 & 1 \\ 1 & \bar{x}_3 & \bar{x}_2 & 0 \\ 0 & 1 & \bar{x}_4 & \bar{x}_3 \\ \bar{x}_4 & 0 & 1 & \bar{x}_1 \end{bmatrix}$$

**(ii)** Let us now consider the HCA <225, 30, 30, 225>PB. As $\partial f_{225}/\partial x_i = \partial f_{30}/\partial x_i\ \forall i \in \{1,2,3\}$ ($\because \partial \bar{f}/\partial x_j = \partial f/\partial x_j\ \forall j$, as established in the proof of theorem 4.2), it is clear that this HCA will have the same Jacobian matrix $J_{30}|_{PB}$ shown in corollary (i). Thus, in general, we can say that if we are given an $n \times n$ Jacobian matrix, which *resembles* that of a UCA of *n* cells, the matrix may actually belong to any one of $2^n$ different CAs, of which only 2 are uniform and the rest are hybrid. In this context, "*resemblance* to the Jacobian matrix of a UCA" means that the vector formed by the diagonal element of each row, along with its two neighbors, *in the correct order*, is *essentially* the same for all the rows (e.g. in $J_{30}|_{PB}$ considered in corollary (i), the relevant vectors are $[1\ \bar{x}_2\ \bar{x}_1]$, $[1\ \bar{x}_3\ \bar{x}_2]$, $[1\ \bar{x}_4\ \bar{x}_3]$, $[1\ \bar{x}_1\ \bar{x}_4]$ which are of the general form $[1\ \overline{rb_i}\ \bar{x}_i]$, $i = 1,2,3,4$, $\overline{rb_i}$ being the right neighbor of $\bar{x}_i$).

For example, <30,30,30,30>PB, <225,225,225,225> PB, <30,225,30,30> PB, <30,225,225> PB, <225,225,30,225> PB are some of the $2^4 = 16$ 4-bit CAs characterized by $J_{30}|_{PB}$.

**(iii)** For a linear CA, whether uniform or hybrid, the Jacobian matrix is a unique **constant binary matrix** [1][4] but *the converse is not true*. This is because the complement of each linear rule is itself necessarily a *non-linear* rule. e.g. The UCA <60,60,60,60>NB, where Rule 60 is a linear rule, is characterized by the Jacobian matrix

$$J_{60}|_{NB} = \begin{bmatrix} 1 & 1 & 0 & 0 \\ 0 & 1 & 1 & 0 \\ 0 & 1 & 1 & 0 \\ 0 & 0 & 1 & 1 \end{bmatrix}.$$

The complement of Rule 60 is Rule 195(=255 – 60); so, the non-linear UCA <195,195,195,195>NB gives the same Jacobian matrix $J_{195}|_{NB} = J_{60}|_{NB}$. However, as the set of *affine functions* comprises the linear rules and their complements, we conclude that *if the Jacobian matrix of a UCA/HCA is constant, all the rules involved must be affine Boolean rules*.

### 3.2. Classification of Boolean rules of 3 variables based on H.D's from the set of affine rules

A Boolean rule of 3 variables is said to belong to **Class *m*** if *m* is the minimum possible H.D. of the non-linear rule from any linear rule of 3 variables, i.e. there exists at least one linear rule such that the H.D. of the rule under consideration from this linear rule is *m* and, if *m'* is the H.D. of the said rule from any other linear rule, then *m'* is larger than or equal to *m*. Any n-bit binary string $x = (x_1, x_2, \ldots x_n)$ divides (or partitions) the entire $2^n$ possible n-bit strings into disjoint equivalence classes such that the strings present in a particular class have the same H.D with $x = (x_1, x_2, \ldots x_n)$. As any binary string of length $2^n$ corresponds to an n-variable Boolean function, therefore any Boolean function g also divides (or partitions) the entire Boolean functions into disjoint equivalence classes.

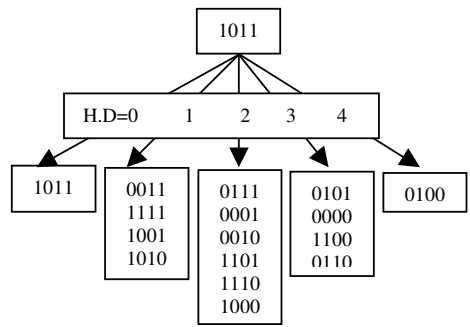

[Fig-3.2.1. Shows a 2-variable Boolean function "1011" classifies all possible 2-variable Boolean functions according to their degree of non-linearity.]

**Observations:**
1. The functions present in a particular class have the same H.D with g.
2. If the H.D between g and any member of a class is k then the cardinality of that class is $^{2^n}c_k$ for $0 \leq k \leq 2^n$.
3. Number of classes possible is $2^n+1$ and can be identified by CLASS 0, CLASS 1, CLASS 2 etc. Here CLASS k means all the members in this class have the H.D k with g for $0 \leq k \leq 2^n$.
4. All the members of CLASS k and CLASS ($2^n$-k) are complement to each other for $0 \leq k \leq 2^n$.

## 4. Importance of the Jacobian matrix in the context of the evolution of a CA

### 4.1 The State Transition Diagram of a CA

The evolution of a CA can be completely described by a diagram in which each state is connected to its successor by a properly directed line-segment. This diagram is called the State Transition Diagram (abbreviated as S.T.D.) of the CA. In other words, the S.T.D. of a CA is essentially a directed graph where each node represents one of the states of the CA and the *edges* signify transitions from one state to another. The S.T.D. of the UCA Rule 170 NB considered is shown below:

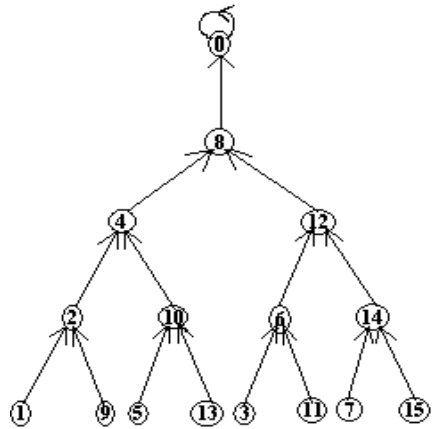

[Fig-4.1. S.T.D. of <170,170,170,170>NB]

### 4.2. The Jacobian matrices of linear and Non-linear CAs

For any linear CA, as already stated, the Jacobian matrix is identically equal to a matrix of '0's and '1's, irrespective of the present state. Moreover, for a linear CA, the following relation holds for any instant $t$ : $(X^{t+1})^T = [F(X^t)]^T = J \cdot (X^t)^T$ where $(X^k)^T$ denotes the transpose of the $1 \times n$ row-vector $X^k$, $k = t, t+1$. Henceforth, for the sake of convenience, the superscript 'T' will be dropped, whenever this does not cause any ambiguity, and the symbol $X^t$ will often be taken to represent the $n \times 1$ column-vector. Similarly for $X^{t+1}$, $F(X^t)$. Thus, $X^{t+1} = F(X^t) = J.X^t$ for a linear CA. Furthermore, for a linear CA, we can not only obtain the successor of each state by simply multiplying its Jacobian matrix with the present state (instead of applying the local mappings to individual cells) but can also deduce all the properties of the State Transition Diagram directly from the algebraic properties (such as rank, nullity, determinant etc.) of the said Jacobian Matrix which is a constant binary matrix; thus the Jacobian Matrix acts as a *linear handle* for the linear CAs. As such, the STDs of linear CAs are predictable and symmetric in structure.
For a non-linear CA, the Jacobian matrix cannot act as linear handle because:
(a) $X^{t+1} \neq J.X^t$ in general
(b) $J$ is, in general, itself a function of $X^t$ so that its matrix properties change depending on the present state $X^t$ of the CA.

### 4.3. Characteristic of non-linear CAs from linear CA

The STD structure of a Nonlinear Rule *g* can be constructed from the nearest linear rule *f* by detecting the deviant states as follows:
1. Find all the 3-bit input strings for which, bit mismatch occurs in the Truth table output of both *f* and *g*. same thing can also be done by looking at the ANF expression of both *f* and *g*.

2. Search these 3-bit strings (say "$x_1y_1z_1$") in the space of all possible $2^n$ states and the states (n-bit strings) containing the string "$x_1y_1z_1$" are all deviant states. Searching should be done in the sense of both Null boundary and Periodic boundary. This can also be done by starting from the decimal value of "$x_1y_1z_1$"=p as a root and then constructing a binary tree of children 2p and 2p+1 extending up to $2^n$-1. All the nodes in this binary tree will represent the deviant states.
3. Only the successor of these deviant states is to be computed in the STD of *g* and for all other non-deviant states the STD's are same with *f* i.e. for all possible deviant state u the edges (u, v) are deleted from the STD of *f* and a new state v' is computed such that (u, v') becomes an edge in the STD of *g*. v' can be computed from v by complementing the bit positions for which the neighborhood string in u is "$x_1y_1z_1$". This can also be done directly by adding or subtracting $2^q$ to the decimal value of v depending on the middle bit position, i.e. $y_1$ as 0 or 1 respectively where $y_1$ is the $(q+1)^{th}$ bit position from right.

**Illustration:** Consider an arbitrary Non-linear rule say Rule 218 and its nearest linear rule is Rule 90. Table 4.1 shows that these two rules are different from each other in their MSB position i.e. for the string "111". So for 4-bit CA the deviant states are 7=(0 **1 1 1**)$_2$, 14=(1 **1 1** 0)$_2$ and 15=(1 **1 1 1**)$_2$ and others are non-deviant states. It can be observed in both the graphs that the successors for all non-deviant states are same but for deviant states they are different. Therefore for 13 non-deviant states the Jacobian matrix for the linear rule: Rule 90 can serve as a handle for the non-linear rule 218. For 3 deviant states, 3 other matrices can be constructed by changing one or more row values in the Jacobian matrix of Rule 90. Fig 4.3.2 shows that $3^{rd}$ and $2^{nd}$ rows have to be changed for deviant states 7 and 14 respectively and for the deviant state 15 two row values both $2^{nd}$ and $3^{rd}$ has to be changed.

| Dec. Value | X | Y | Z | Rule 218 | Rule 90 |
|---|---|---|---|---|---|
| 0 | 0 | 0 | 0 | 0 | 0 |
| 1 | 0 | 0 | 1 | 1 | 1 |
| 2 | 0 | 1 | 0 | 0 | 0 |
| 3 | 0 | 1 | 1 | 1 | 1 |
| 4 | 1 | 0 | 0 | 1 | 1 |
| 5 | 1 | 0 | 1 | 0 | 0 |
| 6 | 1 | 1 | 0 | 1 | 1 |
| **7** | **1** | **1** | **1** | **1** | **0** |

[Table 4.1: Shows Rule 218 and Rule 90 are different from each other in their MSB position i.e. for the string "111".]

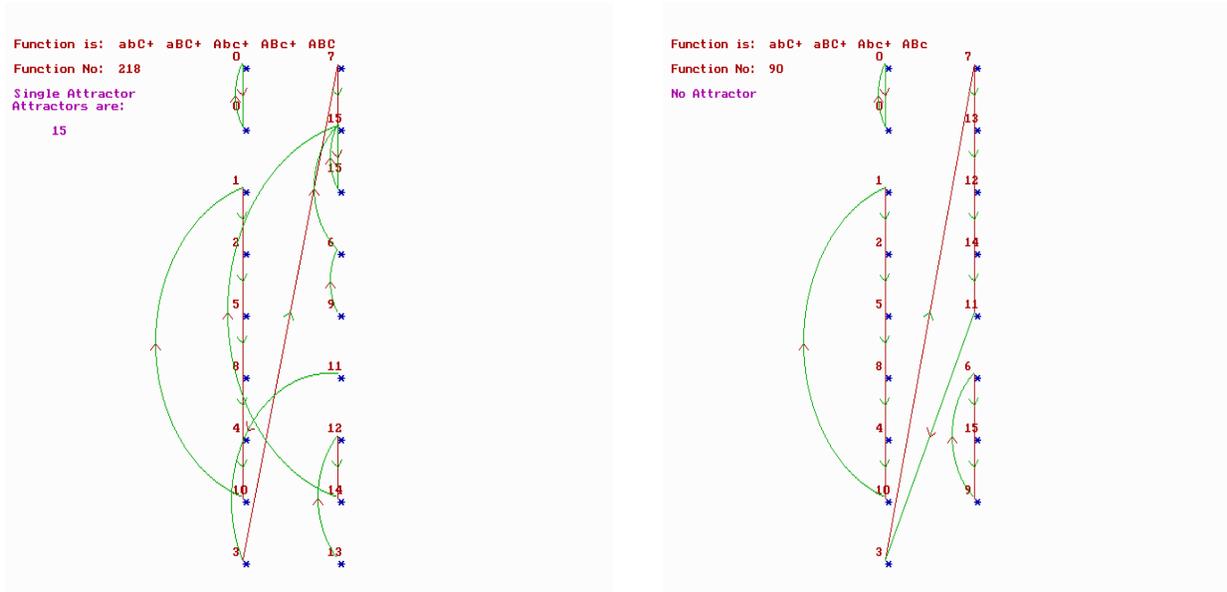

[Fig-4.3.1 S.T.D. of <218, 218, 218, 218>NB and <90, 90, 90, 90>NB]

$$\begin{pmatrix} 0 & 1 & 0 & 0 \\ 1 & 1 & 0 & 0 \\ 0 & 0 & 1 & 0 \\ 0 & 0 & 1 & 1 \end{pmatrix} \text{for state 7,} \begin{pmatrix} 0 & 1 & 0 & 0 \\ 0 & 1 & 0 & 0 \\ 0 & 1 & 1 & 0 \\ 0 & 0 & 1 & 1 \end{pmatrix} \text{for state 14,} \begin{pmatrix} 0 & 1 & 0 & 0 \\ 0 & 1 & 0 & 0 \\ 0 & 0 & 1 & 0 \\ 0 & 0 & 1 & 1 \end{pmatrix} \text{for state 15.}$$

$$J_{90} = \begin{pmatrix} 0 & 1 & 0 & 0 \\ 1 & 1 & 0 & 0 \\ 0 & 1 & 1 & 0 \\ 0 & 0 & 1 & 1 \end{pmatrix} \text{for all the non-deviant states } 1,2,3,4,5,6,8,9,10,11,12,13.$$

[Fig-4.3.2 Matrices for the non-linear rule: Rule 218]

**Observation:**

1. The Jacobean matrix for all the non-deviant states for the non-linear rule *g* is exactly same as the Jacobean matrix for the nearest linear rule *f*. For all the deviant states u, the transformation matrix for the rule *g* can be obtained by changing some row values in the Jacobean matrix of rule *f*. Again which rows are to be changed depends upon the cell position of u for which the neighborhood string is "$x_1y_1z_1$". This means all the states can be divided into two classes: A class of deviant states and a class of non-deviant states. One Jacobian matrix (for the nearest linear rule *f*) can be used for the class of all non-deviant states where as k other matrices are used for the class of all k-deviant states on assuming number of deviant states is k.
2. For a given n-bit string and a non-linear function *g* if k (number of deviant states) goes on increasing definitely the non-linearity is going to increase. So this could be a good measure of non-linearity. Now one can define a measure as a ratio of k and $2^n$(total number of states). Note that this ratio is a fraction, when it is 0 implying no deviant states and the function is linear on the other hand when it is 1 the function is maximally non-linear.
3. The transformation matrix for a deviant state u gives, a single edge (u, v) in the STD for the non-linear rule *g*. Because other matrices are used for other edges in the STD, therefore it is not possible to predict other edges or other state information (like predecessors, successors, non-reachable states etc) using this local transformation matrix at u. But this information can be obtained in linear STD's because a single Jacobian matrix exists for all the states. One think can be said here is that some information can be extracted for all the non-deviant states by the rule matrix related with the nearest linear rule. Collectively the matrices as associated with any arbitrary non-linear rules may throw some more light in this area in terms of some neat and novel formulation.

It should be noted that as the length of CA increases the number of deviant state increases exponentially and thus number of linear handles also will go on increase exponentially. So to obtain a minimal set of matrices for all the deviant states is still an open optimization problem. Our next section throws some light to get a minimal set of matrices using the ANF of an arbitrary CA rule.

## 5. Deriving a set of matrices for an arbitrary one dimensional CA

### 5.1 Mathematical Observations

In this section we will discuss about the Boolean functions of 3 variables because at first we are concentrating on One-dimensional CA. Here we proposed a technique by which every Non-linear rule in 3-variables can be characterized by a sequence of matrices applied in an arbitrary n-bit string except the trivial null string. We have given a proof that for the null string and odd numbered rules a matrix like handle cannot be possible but this can be handled using bit wise complement method.

As "$xyz \oplus xy \oplus xz \oplus x \oplus yz \oplus y \oplus z \oplus 1$" is the ANF for the rule $f_1$ in 3-variables Hence, the rules whose ANF is $\{f_{xyz}, f_{xy}, f_{xz}, f_x, f_{yz}, f_y, f_z, f_1\}$ can be treated as the fundamental rules and other rules can be generated from these. Out of these rules $\{f_x, f_y, f_z\}$ are linear, rest 5 rules are Non-linear.

Our above discussion says that, for every linear rule, we have a single matrix for all $2^n$ possible n-bit strings. But, for Non-linear rules, one matrix is not sufficient for all $2^n$ possible n-bit strings. To get an illustration of this, let us focus our attention in the following theorems.

**Theorem-5.1:** No matrix exists for Rule 1 for the trivial n-bit null string '000…0'.

**Proof:** Consider the trivial input string of length 'n' where all bits are 0. Apply Rule 1= $f_1$ to it. So output strings are always an n-bit string with all bits are 1's.

$$(0\ 0\ 0\ 0\ .\ .\ .\ 0) \xrightarrow{Rule\ 1} (1\ 1\ 1\ 1\ .\ .\ .\ 1)$$

Matrix for this kind of input can not be possible because when we multiply any matrix of order $(n \times n)$ with the null string (0000…0) and equate it with the output (1111…1) this gives us an invalid identity 0=1, which is a contradiction.

As a result of the theorem 5.1, we get the following theorem.

**Theorem-5.2:** 128, odd number rules are there for which a linear handle like matrix can not be constructed for the trivial input string 0=000…. 0, n-number of 0's.

**Proof:** From theorem 5.1, we know that for Rule 1 no matrix can be constructed particularly for null string because Rule 1 produces an output where all bits are 1. Again addition of any fundamental rule with Rule 1 also produces an output where all bits are 1. These rules are 128 in number. Clearly all these rules are Non-linear. If we will use the wolfram naming convention then these are all odd number rules. Hence proved.

From theorem-5.1 and theorem-5.2 it is clear that, for 128 odd numbered rules no matrix exist which when multiplied with null string gives the corresponding n bit output where all values are 1 but the complement of each bit after multiplication gives the output. So for all these odd number rules and input null string the formula used for getting the output is $y = (Ax)^C$ Where $x = (0,0,...,0)$ and $y = (1,1,...,1)$ and $A$ is any matrix of dimension $(n \times n)$.

**Theorem 5.3:** For any input n-bit string u except the trivial string, one can construct $2^{n^2-n}$ possible matrices for a particular rule, whether that rule may be linear or Non-linear.

**Proof:** Consider an input string of length n. suppose k number of zeros are present in that string where k<n. Then that string must contain (n-k) number of 1's. Without loss of generality we can think of that 1st k-bits are '0' and rest 'n-k' bits are 1. Let the Rule matrix be $(a_{ij})$ for i=1, 2, …n and j=1,2, …n. So after multiplication of this Rule matrix with the input string where first k elements are '0' we get the system of equation as follows:

$$a_{1,k+1} \oplus a_{1,k+2} \oplus \ldots \ldots \oplus a_{1,n} = 0\ or\ 1$$
$$a_{2,k+1} \oplus a_{2,k+2} \oplus \ldots \ldots \oplus a_{2,n} = 0\ or\ 1$$
$$.$$
$$.$$
$$.$$
$$a_{n,k+1} \oplus a_{n,k+2} \oplus \ldots \ldots \oplus a_{n,n} = 0\ or\ 1$$

And the variables $a_{1,j}$ for $j \leq k$ does not play any role in this equation. So, these variables can be assigned either 0 or 1 in $2^k$ ways. Consider the first equation from the above system of n equations. It contains (n-k) variables and each variable can take two values either 0 or 1. Hence total possibilities = $2^{n-k}$. But, as the output is fixed for a particular rule i.e. either 0 or 1. Hence, the number of possibilities = $\frac{2^{n-k}}{2} = 2^{n-k-1}$ satisfies that particular equation. The variables which are not involved in this equation their numbers are k. Those variables can take either 0 or 1 and they are $2^k$ in number. So, $2^{n-k-1} \times 2^k$ many solutions exist for the first equation. Therefore the total solution for the system with n-equations $= \left(2^{n-k-1} \times 2^k\right)^n = \left(2^{n-1}\right)^n = 2^{n^2-n}$. Hence proved.

The above theorem says, for an arbitrary n-bit input, $2^{n^2-n}$ number of matrices can be constructed. So, how to construct those matrices? Observation says that the structure of matrices depends both on the input and the output string. Hence by looking the relationship between the input and output strings for a particular rule algebraically it is possible to find out the structure of that Rule matrix.

Again if two rules Rule X and Rule Y applied to a particular n-bit input string give the same output then the set of rule matrices for both the rules are exactly same. Therefore, the set $\{M_1, M_2, M_3, ..., M_{2^{n^2-n}}\}$ is same for both Rule X and Rule Y if they produce the same output as shown in the following figure:

$$(0 \; 0 \; 1 \; 0 \; . \; . \; . \; 1) \xrightarrow{RuleX \; or \; RuleY} (0 \; 1 \; 0 \; 1 \; . \; . \; . \; 0)$$

**Theorem 5.4:** Using maximum 2 row values, one can construct a transformation matrix of an n-bit input state u.

**Proof:** As the input string u is fixed, the rows corresponding to the cell value 1(or 0) in v (the successor of u) are same. This means that for all 0 values in v the corresponding rows in the transformation matrix could be made same. Similar is the case for all 1 values in v.

The above discussion helps us for calculating the minimal set of matrices for Non-linear rules. In Section-2, the minimal set for a linear rule is a single matrix. But for Non-linear rules a single matrix is not sufficient for all possible n-bit input. Our next section is meant for computing the minimal set for Non-linear rules.

## 5.2 Set of matrices for 4-bit strings

From algebraic manipulation one can easily compute the minimal set of matrices for all possible 4-bit input string only for fundamental rules. It can be noted that this set of matrices may not be unique. Other minimal set of matrices can be possible. Just for illustration here we are presenting the technique using which we have calculated a set of matrices for all the rules and for 4-bit strings.

| Dec.value | input | Rule192 / $f_{xy}$ | output | : | Dec.value | input | Rule192 / $f_{xy}$ | output |
|---|---|---|---|---|---|---|---|---|
| 0 | 0000 | → | 0000 | : | 8 | 1000 | → | 0000 |
| 1 | 0001 | → | 0000 | : | 9 | 1001 | → | 1000 |
| 2 | 0010 | → | 0000 | : | 10 | 1010 | → | 0000 |
| 3 | 0011 | → | 0001 | : | 11 | 1011 | → | 1001 |
| 4 | 0100 | → | 0000 | : | 12 | 1100 | → | 0100 |
| 5 | 0101 | → | 0000 | : | 13 | 1101 | → | 1100 |
| 6 | 0110 | → | 0010 | : | 14 | 1110 | → | 0110 |
| 7 | 0111 | → | 0011 | : | 15 | 1111 | → | 1111 |

It can be observed that the output for the inputs 0, 1, 2, 4, 5, 8, 10 is same that is '0000' As we know a null matrix can transfer every input to zero matrix therefore for those strings null matrix $(0)_{4\times 4}$ is one choice. Similarly the output of 15 i.e. '1111' is again the same '1111' for this identity matrix $(I)_{4\times 4}$ may be one choice. Now for others let the matrix be $(a_{ij})_{4\times 4}$. Now our aim is to find out the values of all $a_{ij}$ where i, j = 1, 2, 3, 4 for which the above matrix which when multiplied by the 4–bit input gives the corresponding output. So multiplying the above matrix by the inputs 3, 6, 7, 9, 11, 12, 13, 14 respectively and equating it to its corresponding outputs one can get the system of equations with some $a_{ij}$ as the variables. Then solving these equations we can obtain two new matrices where $a_{ij}$'s are either 0 or 1. In this way we have computed the minimal set of matrices for 8-fundamental rules and are given below. The corresponding input strings are given in table-1: Where 'I' means identity matrix, 'L' stands for the matrix $f_x$, 'U' stands for the matrix $f_z$, '0' stands for Null matrix, '1' and '2' stands for the first and second matrix in the corresponding set from left to right omitting the matrix '0' and 'I'.

$$f_{xyz} = \left\{ \begin{pmatrix} 1 & 1 & 0 & 1 \\ 1 & 1 & 1 & 0 \\ 0 & 1 & 1 & 1 \\ 1 & 0 & 1 & 1 \end{pmatrix}, (0)_{4\times 4}, (I)_{4\times 4} \right\}, \; f_{xy} = \left\{ \begin{pmatrix} 1 & 0 & 0 & 0 \\ 0 & 0 & 0 & 0 \\ 0 & 1 & 0 & 0 \\ 0 & 0 & 0 & 1 \end{pmatrix}, (0)_{4\times 4}, (I)_{4\times 4}, \begin{pmatrix} 0 & 0 & 0 & 1 \\ 0 & 1 & 0 & 0 \\ 0 & 0 & 1 & 0 \\ 0 & 0 & 0 & 0 \end{pmatrix} \right\}$$

$$f_{xz} = \left\{ \begin{pmatrix} 1 & 0 & 1 & 0 \\ 0 & 0 & 1 & 0 \\ 1 & 0 & 1 & 0 \\ 0 & 0 & 1 & 0 \end{pmatrix}, (0)_{4\times 4}, \begin{pmatrix} 1 & 0 & 1 & 0 \\ 0 & 0 & 0 & 0 \\ 0 & 1 & 0 & 0 \\ 0 & 0 & 0 & 0 \end{pmatrix} \right\}, \; f_x = \left\{ \begin{pmatrix} 0 & 0 & 0 & 1 \\ 1 & 0 & 0 & 0 \\ 0 & 1 & 0 & 0 \\ 0 & 0 & 1 & 0 \end{pmatrix} \right\}$$

$$f_{yz} = \left\{ \begin{pmatrix} 0 & 1 & 0 & 0 \\ 0 & 0 & 0 & 0 \\ 0 & 0 & 1 & 0 \\ 1 & 0 & 0 & 0 \end{pmatrix}, (0)_{4\times 4}, (I)_{4\times 4}, \begin{pmatrix} 1 & 0 & 0 & 0 \\ 0 & 0 & 1 & 0 \\ 0 & 0 & 0 & 1 \\ 0 & 0 & 0 & 0 \end{pmatrix} \right\}, \; f_y = \left\{ (I)_{4\times 4} \right\}$$

$$f_z = \left\{ \begin{pmatrix} 0 & 1 & 0 & 0 \\ 0 & 0 & 1 & 0 \\ 0 & 0 & 0 & 1 \\ 1 & 0 & 0 & 0 \end{pmatrix} \right\}, \; f_1 = \left\{ \begin{pmatrix} 1 & 0 & 1 & 0 \\ 1 & 0 & 1 & 0 \\ 1 & 0 & 1 & 0 \\ 1 & 0 & 1 & 0 \end{pmatrix}, (I)_{4\times 4}, \begin{pmatrix} 0 & 1 & 0 & 1 \\ 0 & 1 & 0 & 1 \\ 0 & 1 & 0 & 1 \\ 0 & 1 & 0 & 1 \end{pmatrix} \right\}$$

| Rules/ Inputs | $f_{xyz}$ | $f_{xy}$ | $f_{xz}$ | $f_x$ | $f_{yz}$ | $f_y$ | $f_z$ | $f_1$ |
|---|---|---|---|---|---|---|---|---|
| 0000 | 0 | 0 | 0 | L | 0 | I | U |   |
| 0001 | 0 | 0 | 0 | L | 0 | I | U | 2 |
| 0010 | 0 | 0 | 0 | L | 0 | I | U | 1 |
| 0011 | 0 | 1 | 0 | L | 1 | I | U | 2 |
| 0100 | 0 | 0 | 0 | L | 0 | I | U | 2 |
| 0101 | 0 | 0 | 2 | L | 0 | I | U | 2 |
| 0110 | 0 | 1 | 0 | L | 2 | I | U | 1, 2 |
| 0111 | 1 | 1 | 2 | L | 2 | I | U | 1 |
| 1000 | 0 | 0 | 0 | L | 0 | I | U | 1 |
| 1001 | 0 | 2 | 0 | L | 1 | I | U | 1, 2 |
| 1010 | 0 | 0 | 1 | L | 0 | I | U | 1 |
| 1011 | 1 | 1 | 1 | L | 1 | I | U | 2 |
| 1100 | 0 | 2 | 0 | L | 2 | I | U | 1 |
| 1101 | 1 | 2 | 2 | L | 1 | I | U | 1 |
| 1110 | 0 | 2 | 1 | L | 2 | I | U | 2 |
| 1111 | I | I | I | L | I | I | U | I |

[Table 5.2.1: shows the entire 4-bit input sting, 8-fundamental rules and their corresponding matrices.]

**5.3 Technique for finding the rule matrix for any arbitrary rule and arbitrary n-bit string:**

In Section 5.2, for each rule (linear or Non-linear) and for each 4-bit string we can have a set of matrices. This section gives a linear time algorithm by which for any rule and for any n-bit string (n>4) a series of matrices can be chosen from the above minimal set. This series of matrices when multiplied with the input vectors using the algorithm given below gives the desired output. The way algorithm is designed is suitable for parallel implementation that further reduces the cost of computation. Another advantage of this algorithm is its space complexity, which is $O(1)$.

**Algorithm:**

Non-linear Rule matrix ( $f_w$ )

    1. Input an n-bit string

    2. Partition the n-bit string into n/2 smaller sub-strings each having length 2.

    3. for i= 1 to n/2

    4. {

    5. if( i=1)

    6.   Choose a 4-bit string say $X_i$ from the location n, 1, 2, 3

7. else if (i=n/2)
8.    Choose a 4-bit string say $X_i$ from the location n-2, n-1, n, 1
9. else
10.    {
11.    Choose 4-bit string say $X_i$ from the location 2(i-1), 2i-1, 2i, 2i+1
12.    Compute a rule matrix $(A_i)_{4\times4}$ for 4-bit string $(X_i)_{4\times1}$ from the minimal set of matrices.
13.    }
14. if $(X_i)_{4\times1} \neq (0,0,0,0)$ then
15.    Compute $A_i X_i = b_i$
16. else if $(X_i)_{4\times1} = (0,0,0,0)$
17.    Compute $(A_i X_i)^T = b_i$
18. Remove two padding bits from $b_i$ and collect the middle two bits
19. }

The minimal set of matrices for each 4-bit input string and for each rule can be computed from 8 fundamental rules.

**Illustration:** Given, an 8-bit string

$$(1\ 0 : 1\ 0 : 1\ 0 : 1\ 1) \xrightarrow{Rule128\ or\ f_{xyz}} (0\ 0 : 0\ 0 : 0\ 0 : 0\ 1)$$
            Input                                                         output

$$1101 \longrightarrow \begin{pmatrix} 1 & 1 & 0 & 1 \\ 1 & 1 & 1 & 0 \\ 0 & 1 & 1 & 1 \\ 1 & 0 & 1 & 1 \end{pmatrix},\ 0101 \longrightarrow \begin{pmatrix} 0 & 0 & 0 & 0 \\ 0 & 0 & 0 & 0 \\ 0 & 0 & 0 & 0 \\ 0 & 0 & 0 & 0 \end{pmatrix}$$

$$0101 \longrightarrow \begin{pmatrix} 0 & 0 & 0 & 0 \\ 0 & 0 & 0 & 0 \\ 0 & 0 & 0 & 0 \\ 0 & 0 & 0 & 0 \end{pmatrix},\ 0111 \longrightarrow \begin{pmatrix} 1 & 1 & 0 & 1 \\ 1 & 1 & 1 & 0 \\ 0 & 1 & 1 & 1 \\ 1 & 0 & 1 & 1 \end{pmatrix}$$

Then, multiply the corresponding matrices with the input 4-bit vectors and neglecting the padding bits we'll get the following output vectors.

$$\begin{pmatrix} 1 & 1 & 0 & 1 \\ 1 & 1 & 1 & 0 \\ 0 & 1 & 1 & 1 \\ 1 & 0 & 1 & 1 \end{pmatrix}\begin{pmatrix} 1 \\ 1 \\ 0 \\ 1 \end{pmatrix} = \begin{pmatrix} 1 \\ 0 \\ 0 \\ 0 \end{pmatrix} \longrightarrow 0\ 0,\ \begin{pmatrix} 0 & 0 & 0 & 0 \\ 0 & 0 & 0 & 0 \\ 0 & 0 & 0 & 0 \\ 0 & 0 & 0 & 0 \end{pmatrix}\begin{pmatrix} 0 \\ 1 \\ 0 \\ 1 \end{pmatrix} = \begin{pmatrix} 0 \\ 0 \\ 0 \\ 0 \end{pmatrix} \longrightarrow 0\ 0$$

$$\begin{pmatrix} 0 & 0 & 0 & 0 \\ 0 & 0 & 0 & 0 \\ 0 & 0 & 0 & 0 \\ 0 & 0 & 0 & 0 \end{pmatrix}\begin{pmatrix} 0 \\ 1 \\ 0 \\ 1 \end{pmatrix} = \begin{pmatrix} 0 \\ 0 \\ 0 \\ 0 \end{pmatrix} \longrightarrow 0\ 0,\ \begin{pmatrix} 1 & 1 & 0 & 1 \\ 1 & 1 & 1 & 0 \\ 0 & 1 & 1 & 1 \\ 1 & 0 & 1 & 1 \end{pmatrix}\begin{pmatrix} 0 \\ 1 \\ 1 \\ 1 \end{pmatrix} = \begin{pmatrix} 0 \\ 0 \\ 1 \\ 0 \end{pmatrix} \longrightarrow 0\ 1$$

Output string: "0 0 0 0 0 0 0 1"

Removing the padding two-bit string from each output string we'll get the final output string as shown in the above figure.

**Efficiency of the proposed Algorithm:**

If the length of the input string is 'n' then the number of partitioned matrices is n/2 and hence n/2 sequence of matrices is required for finding the output and hence the time required to find out all the matrices is $O(n)$. The space complexity is to construct the look up table 5.2.1 which contains (16×8)-1=127, $(4\times 4)$ matrices. Hence it is $O(1)$.

Although the time is linear still one can reduce the constant associated in the order notation by computing algebraically the minimal set matrices of higher bit strings. Also the algorithm is suitable for parallel implementation by assigning each part of 4-bit string to different processors to do the same job.

## 6. Conclusion and future efforts

This paper characterizes an arbitrary uniform one-dimensional Non-linear CA with the help of matrices. Two different approaches have been made. First the matrices can be computed from the STD's of linear CA using deviant and non-deviant states. Second, a complexity of constant space and linear time algorithm is proposed based on which it was shown that Non-linear CA could be captured through linear handles like matrices. Here our study is restricted to 3-neghbourhood CA, but the concept can be easily extended to arbitrary k-neighborhood CA in one-dimension. Although the procedure discussed here gives a sequence of matrices for arbitrary Non-linear CA, but to obtain a minimal set of matrices is still an open optimization problem. It is the firm conviction of the authors that the representative set of linear matrices corresponding to any non-linear CA in one and higher dimensions could lead a long way in non-linear dynamics.